\documentclass{elsart}
\usepackage{natbib}
\begin{document}
\runauthor{Grupe, Leighly, Thomas, and Laurent-Muehleisen}
\begin{frontmatter}
\title{The enigmatic soft X-ray NLS1 \\
RX J0134.2--4258
}
\author[dirk]{Dirk Grupe\thanksref{email}}
\author[karen]{Karen M. Leighly}
\author[hct]{Hans-Christoph Thomas}
\author[sally]{Sally A. Laurent-Muehleisen}

\address[dirk]{Max-Planck-Institut f\"ur extraterrestrische Physik,
Giessenbachstr., D-85748 Garching, Germany}
\thanks[email]{E-mail: dgrupe@mpe.mpg.de}
\address[karen]{Columbia Astrophysics Lab., 550 West 120th St., New York,
NY 10027, USA}
\address[hct]{MPI F\"ur Astrophysik, karl-Schwarzschild-Str. 1, D-85748
Garching, Germany}
\address[sally]{UC-Davis and IGPP/LLNL, 7000 East Av. Livermore, CA 94550, USA}

\begin{abstract}
We report the discovery and analysis of the follow-up {\it ROSAT} pointed
observation, an {\it ASCA} observation and optical and radio observations of
the enigmatic Narrow-Line Seyfert 1 galaxy RX J0134.2--4258.  While
its spectrum was one of the softest observed from an AGN during the
{\it ROSAT} All-Sky Survey, its spectrum was found to be dramatically harder
during a pointed observation although the count rate remained
constant. In the pointed observation we found that the spectrum is
softer when it is fainter, and spectral fitting demonstrates that it is
the hard component that is variable.  The {\it ASCA} observation confirms
the presence of a hard X-ray power law, the slope of which is rather
flat compared with other NLS1s.  Survey and follow-up radio
observations reveal that RX J0134.2--4258 is also unusual in that it
is a member of the rare class of radio-loud NLS1s, and, with R=71, it
holds the current record for the largest radio-to-optical ratio in NLS1s.
We discuss possible scenarios to explain its strange behaviour.
\end{abstract}
\end{frontmatter}

\section{Introduction}

One of the most interesting results from {\it ROSAT} observations of
AGN was the discovery of several X-ray transient AGN.  These objects
were detected to be bright soft X-ray sources during the {\it ROSAT} All Sky
Survey (RASS), but follow-up observations revealed large amplitude
decreases in flux by up to two orders of magnitude.  RX~J0134.2$-$4258
also exhibited X-ray transience, but the behavior was somewhat
different.  During the RASS, it was found to have one of the steepest
spectra exhibited by an AGN. During the follow-up pointed observation
about two years later,
the spectrum was much flatter.  
We report the results of a study of this object using {\it ROSAT}, {\it ASCA}, 
optical and radio observations; details can be found in Grupe et al.\ (2000, A\&A, in press).

\section{Observations}

RX~J0134.2$-$4258 was discovered during the RASS.  Optical
spectroscopic observations showed that the object at the center of the 
RASS error circle is a NLS1 at a redshift of 0.237. 

Based on a small sample of 15 objects, Ulvestad et al.\ 1995 
reported that NLS1s are generally weak radio sources. As discussed by
E.\ Moran in these proceedings, this is not always the case.  We
discovered that RX~J0134.2$-$4258 is a moderately strong radio source.
In the Parkes-MIT-NRAO (PMN) radio catalog its flux density is $55\pm
9$ mJy at 4.85 GHz, and a VLA observation measures a flux density of
$25\pm 1$ mJy at 8.4 GHz.  We infer that the radio spectral index is
rather steep (1.4) and/or that it is a variable radio source.
Comparison of the 8.4 GHz radio flux with the optical spectrum reveals
that RX~J0134.2$-$4258 is radio-loud: the radio to optical ratio R=71,
and the luminosity is $\log(P)=25.3\rm \, W Hz^{-1}$.

{\bf Spectral Variability Observed by {\it ROSAT}:}
The RASS observation revealed that the X-ray spectrum of 
RX~J0134.2$-$4258 is very steep and practically no photons were
detected above 1~keV.  The inferred photon index was
$\alpha_x=6.9\pm2.9$; poor statistics prevented more complex modeling.
The follow-up pointed observation found that the count rate from
RX~J0134.2$-$4258 had not changed substantially. However, the spectrum
was significantly harder.  A power law model with Galactic absorption
did not describe the spectrum adequately.  While the poor energy
resolution of the {\it ROSAT} PSPC means that we cannot definitively
determine the origin of the spectral complexity, we favor a model with a power
law plus a soft component, where $\alpha_x=1.04\pm 1.22$ and
$kT=14\pm 7\rm \,eV$.

RX~J0134.2$-$4258 was variable during the pointed observation.  The
pointed observation was split into 6 intervals (OBIs).
The third OBI has significantly lower flux, and the hardness ratio
analysis showed that the spectrum was temporarily much softer as well.
The spectral variability on short time scales observed during the
pointed observation seems to be similar to the long time-scale
variability observed between the RASS and the pointed observation; that
is, both are produced by a large amplitude change in the hard X-rays,
while the soft X-rays remained nearly constant.

{\bf {\it ASCA} Observation:}
We observed RX~J0134.2$-$4258 using {\it ASCA} for $\sim 45$ ks.  The object
varied during the observation and the amplitude of variability was
consistent with that of other NLS1s with similar hard X-ray luminosity
(see Leighly 1999a).  No spectral variability was detected.

The {\it ASCA} spectrum was adequately modeled with a power law and Galactic
absorption.  The energy index was $0.80 \pm 0.09$, a rather flat value
compared with other NLS1s (e.g.\ Leighly 1999b).  No evidence for
absorption edges at 0.74 or 0.87 keV, appropriate for ionized oxygen
in a warm absorber, was found.  The optical depth upper limits are not
strongly constraining ($\sim 0.5$); however, a large optical depth
warm absorber is clearly ruled out.  Similarly, no evidence for an
iron line was found, with the EW upper limit for a narrow 6.4 keV (rest
frame) line of 300 eV.

The spectrum from the {\it ROSAT} pointed observation and the {\it ASCA} spectrum
were consistent with each other except for a normalization factor in
the overlapping energy band.  Therefore, we elected to model the
spectra simultaneously.  We found that the spectrum could not be
modeled by a power law alone, but a power law plus soft excess model
fit the spectra well.

{\bf Optical observations:}
An optical spectrum was obtained with the ESO 
1.52m telescope during four nights in September 1995 for a total of
3.25 hours  (see Grupe et al.\ 2000).  The spectrum is typical of a 
Narrow-line Seyfert 1 galaxy with extreme FeII emission, weak [OIII]
and a very blue optical continuum. We measured
FWHM(H$\beta$)=900$\pm$100 and FWHM($\rm [OIII]$)=670$\pm$200 km $\rm
s^{-1}$.  Its FeII/H$\beta$ ratio (12.3) is the largest among our
entire sample of soft X-ray selected AGN (Grupe et al.\ 1999). 

\section{Discussion}

RX~J0134.2$-$4258 showed extreme spectral variability between the RASS
and the {\it ROSAT} pointed observation and also during the pointed
observation.  On both long and short time scales, the soft flux
appeared to be constant, while the hard flux varied with large
amplitude.  Nothing exactly like this has ever been reported before,
although behavior somewhat similar was observed from Mrk~766 whose 
spectral variability during the {\it ROSAT} pointed observations was also
consistent with a non-variable soft component (Leighly et al.\ 1996).
We explore possible origins of the spectral variability below; see
Grupe et al.\ (2000) for details.

{\bf Warm Absorber:} Komossa \& Merrschweinchen (2000) speculate that
the spectral variability between the RASS and pointed observation is
due to a partially ionized (warm absorber) cloud traversing the line
of sight.  We find no evidence for a warm absorber in the {\it ASCA}
spectra, however.  Furthermore, we find that for reasonable
parameters, the cloud scenario cannot explain the rather similar
spectral variability observed on shorter time scales during the
pointed observation, simply because for plausible source sizes and
cloud velocities, the time scale of variability required is rather
longer than observed.

{\bf Corona Loss and Recovery:} The steep spectra observed from NLS1s
imply that there must be energy released in the accretion disk to
power the soft X-ray emission.  Therefore, a hard X-ray emitting
corona of energetic particles need not necessarily be present.  It is
possible that during the RASS, the hard X-ray emitting corona was
absent, and it reappeared before the {\it ROSAT} pointed observation and
the {\it ASCA} observation.  A similar phenomenon may have been observed in
NGC~4051; this object was found to be in a very low state recently, 
when the hard X-ray emitting power law was absent, and the spectrum
was dominated by Compton reflection (Guainazzi et al.\
1998).

{\bf Strengthening of the Radio Component:} We discovered that
RX~J0134.2-4258 is a radio-loud NLS1 and that it has a flat hard X-ray
spectrum compared with other NLS1s.  Radio-loud objects are known to
have flatter hard X-ray spectra and to be more powerful X-ray sources,
possibly because of an extra contribution to the X-ray flux associated
with the radio-emitting component.  Thus, another possible explanation
for the spectral variability is an increase in the component
associated with the radio emission. We note, however, that
RX~J0134.2-4258 is a relatively weak radio source, and that the other
two radio-loud NLS1s observed with {\it ASCA} have canonical steep
hard X-ray spectra.

\end{document}